 \documentclass[pdflatex,sn-nature]{sn-jnl}

\usepackage{longtable}
\usepackage{graphicx}%
\usepackage{multirow}%
\usepackage{amsmath,amssymb,amsfonts}%
\usepackage{amsthm}%
\usepackage{mathrsfs}%
\usepackage[title]{appendix}%
\usepackage{xcolor}%
\usepackage{textcomp}%
\usepackage{manyfoot}%
\usepackage{booktabs}%
\usepackage{algorithm}%
\usepackage{algorithmicx}%
\usepackage{algpseudocode}%
\usepackage{listings}%
\usepackage{aas_macros}%

\theoremstyle{thmstyleone}%
\theoremstyle{thmstyletwo}%

\theoremstyle{thmstylethree}%

\begin{document}

\title[Article Title]{Direct Observational Constraint on Upstream Magnetic Field Amplification Reveals the Particle Acceleration Limit in a Supernova Remnant}


\author*[1]{\fnm{Yuya} \sur{Kawabata}}\email{yuyakawabataastro@gmail.com}

\author[1]{\fnm{Takaaki} \sur{Tanaka}}

\author[2]{\fnm{Hiromasa} \sur{Suzuki}}

\author[1]{\fnm{Tsuyoshi} \sur{Inoue}}

\affil[1]{\orgdiv{Department of Physics}, \orgname{Konan University}, \orgaddress{\street{8-9-1 Okamoto, Higashinada}, \city{Kobe}, \state{Hyogo} \postcode{658-8501}, \country{Japan}}}

\affil[2]{\orgdiv{Faculty of Engineering}, \orgname{University of Miyazaki}, \orgaddress{\street{1-1 Gakuen, Kibanadai Nishi}, \city{Miyazaki}, \state{Miyazaki} \postcode{889-2192}, \country{Japan}}}

\abstract{
The origin of Galactic cosmic rays up to the knee energy (PeV) remains a long-standing open question in astrophysics. 
Supernova remnant shocks are leading candidates for accelerating particles to PeV energies, yet this requires strong magnetic field amplification in the upstream region to confine the highest energy particles. 
Despite its importance, the upstream magnetic field strength has long lacked direct observational constraints. 
Here we report the observation of numerous synchrotron X-ray hotspots extending into the upstream of the supernova remnant RX J1713.7$-$3946 in archival Chandra data. 
These hotspots show spatial and temporal properties consistent with synchrotron emission from secondary electrons produced by hadronic interactions between accelerated protons and dense molecular clumps. 
Using the spatial extent of these upstream hotspots as a physical scale, we estimate the proton diffusion length to be $\sim 1~\mathrm{pc}$. 
This measurement provides a constraint on the upstream magnetic field strength of $\sim 10~\mu\mathrm{G}$, which is consistent with recent simulations of Bell-instability-driven magnetic field amplification. 
Our results indicate that the current amplification is insufficient to reach PeV energies, explaining the observed TeV gamma-ray spectral cutoff and providing a diagnostic for assessing PeVatron capability in stellar shocks.
}

\maketitle

\section*{Introduction}
Supernova remnants (SNRs) are leading candidate sources of Galactic cosmic rays up to the ``knee'' energy of 
$\sim 3\mathrm{~PeV}$. 
The paradigm of diffusive shock acceleration\cite{Bell1978,BlandfordEichler1987} (DSA) provides a framework for this process 
as indeed partly verified by X-ray and gamma-ray observations of SNRs\cite{Koyama1995,Aharonian2004,Ackermann2013}. 
Yet direct observational evidence of SNRs acting as PeV particle accelerators (PeVatrons) is still lacking. 

While recent sub-PeV gamma-ray detections by the Large High Altitude Air Shower Observatory (LHAASO)\cite{LHAASO2022} have extended the observable energy range in the Galaxy\cite{Cao2024}, 
identifying the precise acceleration sites and their physical limits requires an understanding of the magnetic field environment surrounding the shock.

The maximum energy of accelerated particles is governed by the degree of magnetic field amplification. 
In their study of the SNR RX~J1713.7$-$3946\cite{Pfeffermann1997}, Uchiyama et al.\cite{Uchiyama2007} discovered mG-scale magnetic fields through 
the rapid temporal variability of X-ray filaments. 
This amplification, likely driven by the shock's interaction with a clumpy interstellar medium\cite{inoue2012}, 
occurs primarily in the shock downstream. 
However, the ultimate reach of a cosmic-ray factory is determined not only by the downstream turbulence 
but essentially by the upstream magnetic field, which dictates the confinement and pre-acceleration of the highest-energy particles. 
Despite its theoretical importance, particularly in the context of the non-resonant Bell instability\cite{Bell2004}, 
the magnetic field strength in the shock upstream has remained almost entirely unconstrained.

In this Article, we constrain the upstream magnetic field of SNR RX J1713.7$-$3946 by utilizing a population 
of X-ray hotspots\cite{Higurashi2020} that extend well beyond the forward shock. 
Through a systematic analysis of archival Chandra data, we identify these hotspots across the entire remnant and interpret 
them as synchrotron emission from secondary electrons produced by upstream protons. 
By using the spatial distribution of these upstream hotspots as a direct measure of the proton diffusion length, we provide an observational constraint on the upstream magnetic field strength. 
This approach allows us to independently evaluate the PeVatron potential of RX J1713.7$-$3946, bridging the gap between theoretical predictions of the Bell instability and the observed TeV gamma-ray spectral cutoff\cite{Aharonian2006,Aharonian2007,HESSCollaboration2018}.

\section*{Results and Discussion}
\begin{figure}
  \begin{center}
    \includegraphics[width=8.5cm]{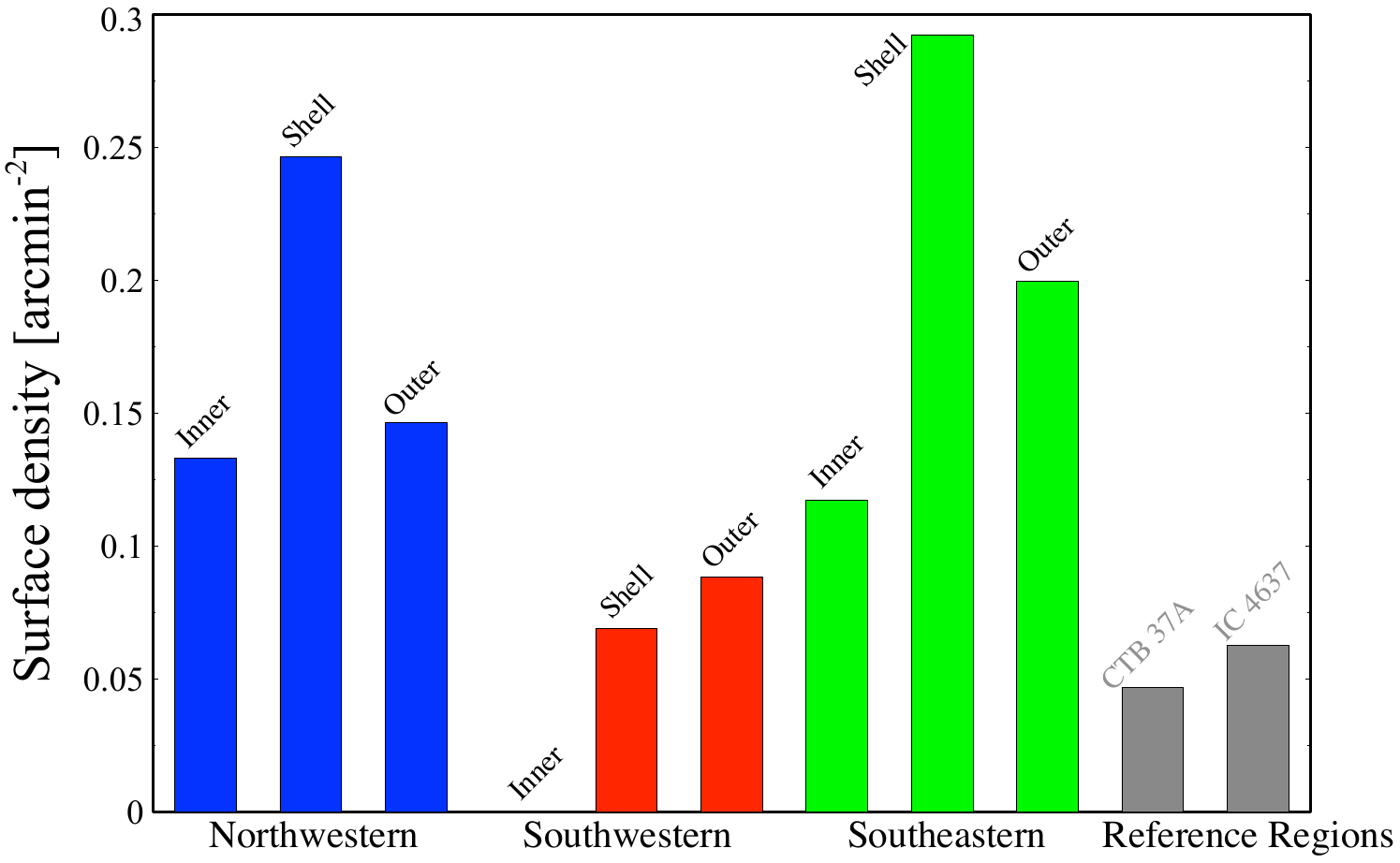}
  \end{center}
  \caption{\textbf{Surface densities of X-ray hotspots in SNR RX~J1713.7$-$3946 and in the reference regions.} 
 The colored bars represent the surface density of X-ray hotspots (number per unit area) identified in the northwestern (NW; blue), southwestern (SW; red), and southeastern (SE; green) regions of SNR RX~J1713.7$-$3946.  Gray bars denote the baseline density measured in reference regions (SNR CTB~37A and planetary nebula IC~4637). The regions are subdivided into ``Inner'', ``Shell'', and ``Outer'' sub-regions relative to the shock front, as defined in Fig.~\ref{fig:NW,SW,SE image}. The hotspot density in RX~J1713.7$-$3946 exceeds the levels measured in the reference regions across nearly all regions, particularly in the ``Outer'' sub-regions of the NW and SE.}\label{fig:SurfaceDensity}
\end{figure}

To investigate the spatial extent of particle acceleration, we performed a systematic search for point-like X-ray sources, or ``hotspots'', 
across the entire extent of RX~J1713.7$-$3946 using archival Chandra data. 
As shown in Fig.~\ref{fig:NW,SW,SE image}, our analysis reveals that hotspot populations are not localized to the northwestern (NW) 
region\cite{Higurashi2020} but are instead distributed across the southwestern (SW) and southeastern (SE) rims.
To determine whether these features are physically associated with the SNR, we compared their surface density with those of reference regions (CTB~37A and IC~4637). 
The results, summarized in Fig.~\ref{fig:SurfaceDensity}, demonstrate that the hotspot density in RX~J1713.7$-$3946 is higher than the level measured in the reference regions across nearly all sub-regions. 
This high density persists even in the ``Outer'' sub-regions located well beyond the forward shock (Fig.~\ref{fig:SurfaceDensity}). 
This spatial coincidence is consistent with the upstream hotspots being driven by physical processes originating from the SNR, such as the diffusion of accelerated protons into the surrounding interstellar medium. 
While the SW region shows a slightly higher potential for contamination, the overall remnant-wide distribution points toward a common mechanism for particle diffusion into the shock upstream.

A plausible explanation for the upstream hotspots is that they are illuminated by high-energy protons in the shock upstream. 
As proposed by Inoue et al.\cite{inoue2012}, dense clumps remain in the shock upstream, where the diffuse inter-cloud gas was previously swept away 
by the stellar wind from the progenitor star before the supernova explosion. 
The highest-energy protons diffuse into the shock upstream and interact with these clumps ($pp$ interactions). 
These hadronic interactions generate $\pi^\pm$, which decay into $\mu^\pm$ and subsequently into $e^\pm$. 
These secondary electrons and positrons then produce synchrotron X-ray emission in the localized magnetic fields associated with the clumps, 
manifesting as the observed X-ray hotspots in the shock upstream. 

\begin{figure}
  \begin{center}
    \includegraphics[width=0.7\linewidth]{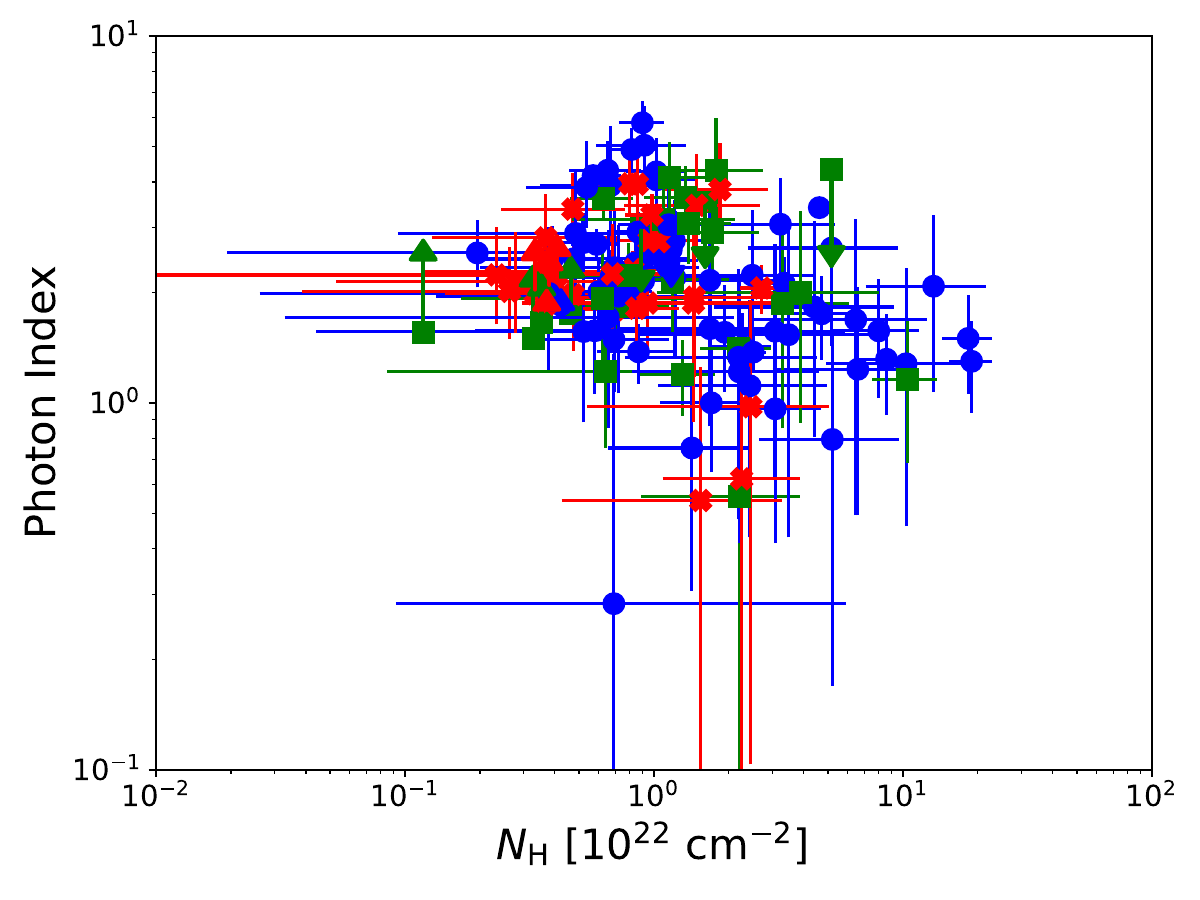}
  \end{center}
  \caption{\textbf{Relation between the hydrogen column density ($N_\mathrm{H}$) and the photon index ($\Gamma$) derived from the spectral fitting.} The red crosses, blue circles, and green squares represent the ``Inner'', ``Shell'', and ``Outer'' sub-regions, respectively. Error bars represent the $1\sigma$ confidence level derived from the simultaneous spectral fitting. 
The data points show a hint of an anti-correlation, which is compatible with the hadronic scenario.}
\label{fig:NW_SW_SE_anticorrelation}
\end{figure}

The proposed scenario is further tested with our spectral analysis, 
where we employed a simultaneous spectral fitting method that self-consistently accounts for both source and background components (see Methods). 
This approach yields smaller statistical uncertainties on the spectral parameters of individual hotspots.  
We find a hint of an anti-correlation between the hydrogen column density ($N_\mathrm{H}$) and the photon index ($\Gamma$) of the hotspots, including those in the shock upstream (Fig.~\ref{fig:NW_SW_SE_anticorrelation}).
Although the anti-correlation may not be significant enough, it can be explained by the mechanism proposed for the downstream hotspots\cite{Higurashi2020}: 
higher-energy protons penetrate more deeply into the dense cores of the molecular clumps. 
Such interactions preferentially enhance the production of higher-energy secondary $e^{\pm}$ in high-density regions, resulting in a harder synchrotron X-ray spectrum (smaller $\Gamma$) in regions with higher column densities ($N_\mathrm{H}$).

Two further properties support the secondary-electron interpretation. 
First, the ratio of the summed X-ray flux of the upstream hotspots to the TeV gamma-ray flux from the same ``Outer'' sub-regions is $F_\mathrm{X}/F_{\gamma} \sim 0.1$, in agreement with the expectation in this scenario (see Methods). 
Second, the upstream hotspots show no significant flux variability. 
In the downstream hotspots associated with $\sim \mathrm{mG}$ molecular clumps whose X-ray variabilities are observed\cite{Higurashi2020}, the synchrotron cooling time of the X-ray-emitting electrons is only $\sim 1~\mathrm{yr}$, and, thus, the hotspot brightness tracks the instantaneous supply of electrons. 
The absence of significant variability of the upstream hotspots over the decade-long baseline (see Methods and Supplementary Fig.~\ref{sfig:variable}), therefore, indicates constant supply of accelerated protons into the upstream in a steady state. 

\begin{figure}
  \begin{center}
    \includegraphics[width=\linewidth]{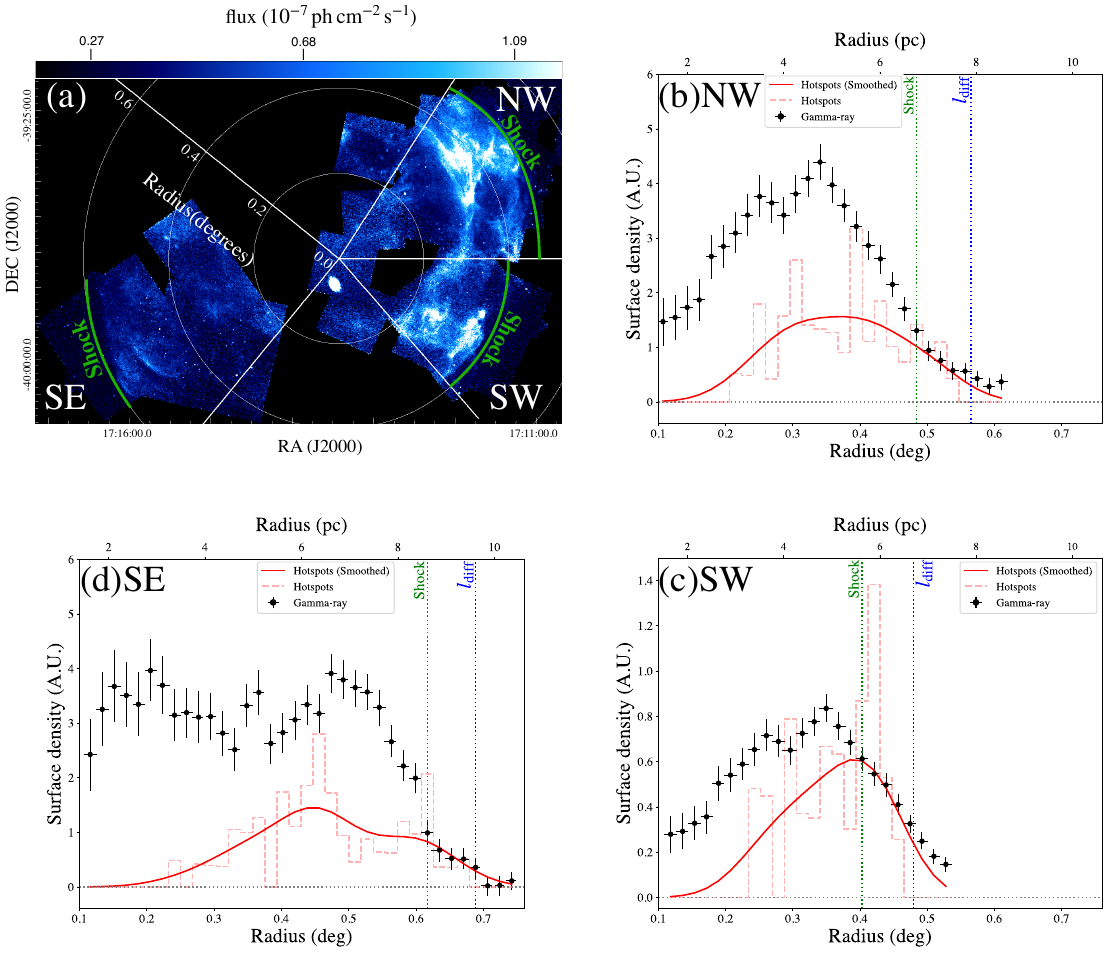}
  \end{center}
  \caption{
  \textbf{Radial profiles of the X-ray hotspot surface density and of the TeV gamma-ray brightness.} The panel (a) is a Chandra X-ray image showing the sectors used for the profile extraction. The panels (b), (c), and (d) are radial profiles of the hotspot surface density as observed (magenta histograms) and after being smoothed with a Gaussian kernel matching the point spread function of H.E.S.S. for the (b) NW ($n=79$), (c) SW ($n=22$), and (d) SE ($n=94$) regions, respectively. 
For comparison, radial profiles of the TeV gamma-ray brightness (black points from H.E.S.S.\cite{HESSCollaboration2018}) are also plotted. 
Error bars on the black points represent the $1\sigma$ statistical uncertainties reported by H.E.S.S.\cite{HESSCollaboration2018}.
Vertical green dashed lines indicate the estimated position of the forward shock determined based on the X-ray image while the blue dashed lines denote the locations corresponding to $l_\mathrm{diff}$ estimated based on the smoothed distributions. The distance to RX~J1713.7$-$3946 was assumed to be $1~\mathrm{kpc}$\cite{Koyama1997,fukui2003} to convert angular distances into physical scales. }
\label{fig:radial_profile}
\end{figure}

The spatial distribution of the upstream hotspots traces the extent to which high-energy protons have reached into the shock upstream. 
In Fig.~\ref{fig:radial_profile}, we compare the radial profile of the hotspot surface density with the TeV gamma-ray emission profile observed by H.E.S.S.\cite{HESSCollaboration2018}. 
The two independent datasets show a similar radial dependence; the hotspot distribution traces the TeV gamma-ray ``tail'' that extends nearly $1~\mathrm{pc}$ beyond the forward shock. 
This is consistent with the X-ray hotspots and the extended TeV emission trace the same population of high-energy particles diffusing into the upstream medium. 
We note that the coincidence constrains the spatial extent of particle diffusion, not the emission mechanism itself; the hadronic interpretation proposed by previous studies\cite{fukui2012,inoue2012,GabiciAharonian2014,fukui2021} remains plausible, and our constraint on the upstream magnetic field holds in either case (see Methods). 
The differences between the two profiles in the ``Shell'' and ``Inner'' sub-regions, by contrast, suggest distinct populations of radiating particles inside the remnant. 

The $\sim 1~\mathrm{pc}$ extent of the upstream hotspots reflects the proton diffusion length rather than the spatial cutoff of the target material. 
The dense molecular gas traced by CO observations\cite{fukui2012} extends well beyond the outermost hotspots, and thus the decline of the hotspot surface density in the upstream cannot be attributed to a lack of target clumps. 
We thus identify the $1/e$ decline scale of the smoothed hotspot surface density ahead of the forward shock with the proton diffusion length, $l_\mathrm{diff} \sim 1~\mathrm{pc}$ (blue dashed lines in Fig.~\ref{fig:radial_profile}; see Methods).

Based on the observed diffusion length, $l_\mathrm{diff} \sim 1~\mathrm{pc}$, we can constrain the upstream magnetic field strength, $B_\mathrm{up}$ (see Methods). 
Even with assuming that the particles are at the limit of their confinement ($\eta = 1$), the measured $l_\mathrm{diff}$ value implies an upstream magnetic field of $B_\mathrm{up} \sim 10~\mu\mathrm{G}$ (Eq.~\ref{formula:Bdiff}). 
This value agrees with recent numerical simulations of the Bell instability\cite{Bell2004, inoue2024}, 
which predict that magnetic field amplification in the precursor of such a high-velocity shock is limited to this level. 
An upstream field of $B_\mathrm{up} \sim 10~\mu\mathrm{G}$ limits the maximum proton energy to $\sim 100~\mathrm{TeV}$ (Eq.~\ref{formula:Emax}) 
for an SNR with an age of $\sim 2000~\mathrm{yr}$ and a shock velocity of $V_\mathrm{sh} \sim 3000~\mathrm{km}~\mathrm{s}^{-1}$\cite{TsujiUchiyama2016,Tanaka2020,Acero2017}. 
This result provides an independent explanation for the spectral cutoff observed in the TeV gamma-ray spectrum\cite{HESSCollaboration2018}, indicating that RX~J1713.7$-$3946 is not currently acting as a PeVatron. 
However, the possibility remains that it functioned as a PeVatron during an earlier stage of its SNR evolution when the shock velocity was higher.

The inferred $l_\mathrm{diff} \sim 1~\mathrm{pc}$ is similar across the NW, SW, and SE regions despite their different shock velocities. 
This convergence is expected if the upstream turbulence saturates near the Bohm limit: regardless of the local shock velocity or ambient density, the amplification cannot exceed this limit, and the proton diffusion length asymptotes to a common value of $\sim 1~\mathrm{pc}$.

\section*{Conclusions}
Our result demonstrates that RX J1713.7$-$3946, with its environment of dense molecular clumps in the shock upstream, provides a laboratory in which to measure the particle diffusion length and the upstream magnetic field. While such hotspots have not, to the best of our knowledge, been clearly identified in other SNRs, our findings establish a physical benchmark for the particle acceleration limit, showing that this SNR currently lacks the magnetic field amplification required to reach PeV energies.

\clearpage

\begin{figure}
  \begin{center}
    \includegraphics[width=0.85\linewidth]{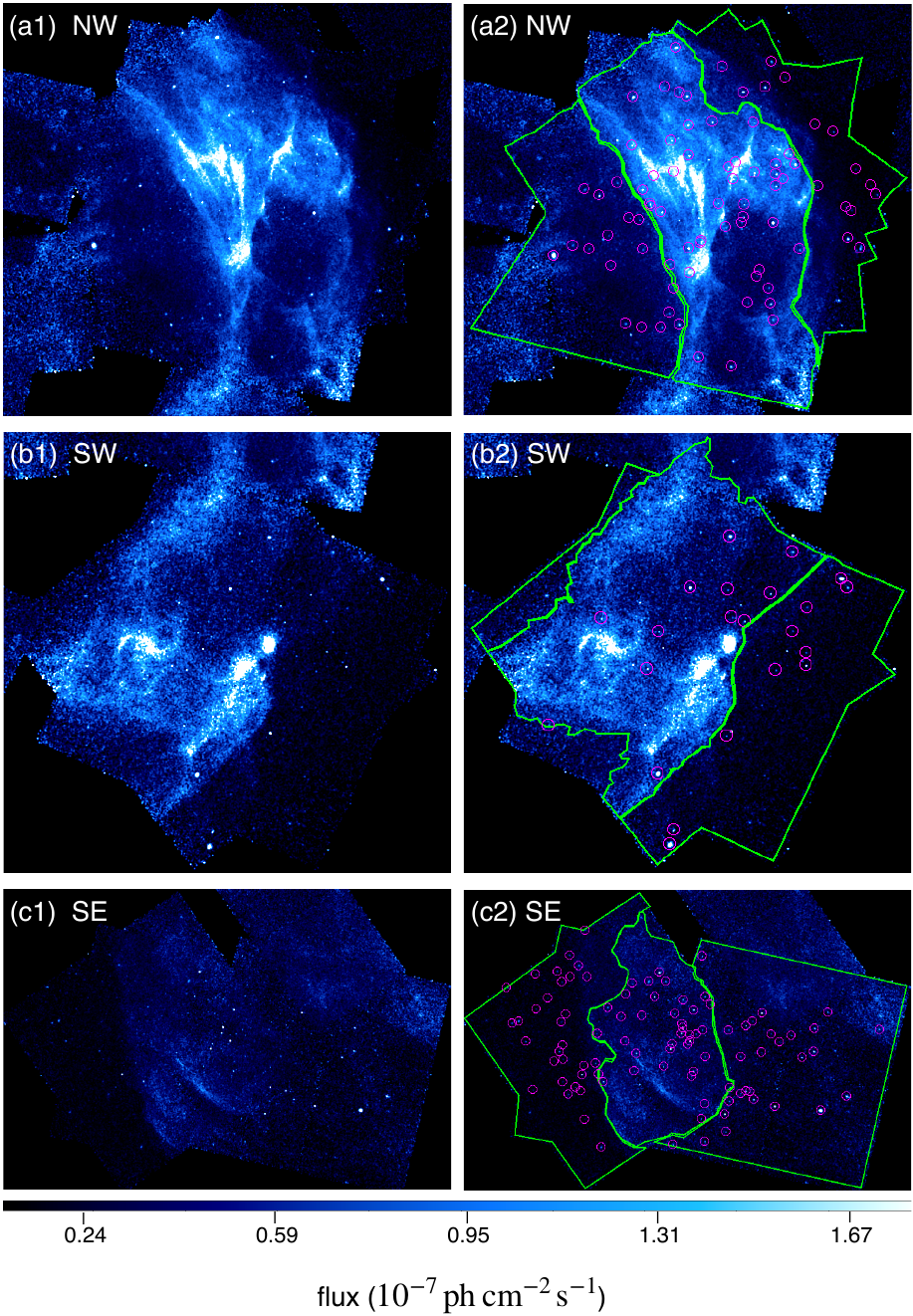}
  \end{center}
  \caption{\textbf{Locations of the X-ray hotspots in SNR RX~J1713.7$-$3946.} The panels (a), (b), and (c) are Chandra X-ray ($0.5\textrm{--}7.0~\mathrm{keV}$) images of the (a1) NW, (b1) SW, and (c1) SE regions, respectively. The panels (a2), (b2), and (c2) are the same but with magenta open circles that denote the locations of point-like X-ray sources (``hotspots'') identified with a statistical significance of $> 3\sigma$; point-like features without circles do not reach this significance threshold (see Methods). The green lines define the sub-regions (``Inner'', ``Shell'', and ``Outer'') categorized based on the X-ray morphology relative to the forward shock. Notably, a substantial population of hotspots is detected in the ``Outer'' sub-regions, which correspond to the shock upstream. These upstream hotspots, distributed universally across the remnant, provide a direct probe into the environment of the shock upstream.}
  \label{fig:NW,SW,SE image}
\end{figure}

\clearpage

\section*{Methods}\label{sec:methods}
\subsection*{Observations and Data Reduction}\label{section:Observations and Data Reduction}
We analyzed archival data from observations of RX~J1713.7$-$3946 performed with the Advanced CCD Imaging Spectrometer (ACIS) aboard the Chandra X-ray Observatory. 
The observations used in this study are summarized in Supplementary Table~\ref{sTable:Chandra-Obs}. 
The observations cover the NW, southwestern (SW), and southeastern (SE) regions of the SNR. 
For the NW observations, we analyzed the same datasets that were used by Higurashi et al.\cite{Higurashi2020}. 
Following Higurashi et al.\cite{Higurashi2020}, we also used observations of CTB~37A and IC~4637 as references.  
All the data were reprocessed using \texttt{chandra$\_$repro} in Chandra Interactive Analysis of Observations (CIAO) version 4.11.0 with CALDB version 4.16.0. 

\subsection*{Detection of X-ray Hotspots}\label{section:Detection}
To search for X-ray hotspots, we employed the \texttt{wavdetect} tool. 
As a result, we detected 147, 50, and 115 point-like sources in the NW, SW, and SE regions, respectively.   
These include most of the 134 sources previously identified in the NW region by Higurashi et al.\cite{Higurashi2020}.
For each detected source, we calculated its significance $S$ in units of standard deviations ($\sigma$) using the equation by Li \& Ma\cite{LiMa1983}: 
\begin{eqnarray}
  S = \sqrt{2}\left\{
    N_\mathrm{on}\,\ln\left[\frac{1+\alpha}{\alpha}
    \left(\frac{N_\mathrm{on}}{N_\mathrm{on}+N_\mathrm{off}}\right)\right]
    \right. 
   + \left. N_\mathrm{off}\,\ln\left[(1+\alpha)
    \left(\frac{N_\mathrm{off}}{N_\mathrm{on}+N_\mathrm{off}}\right)\right]
    \right\}^{1/2}, 
  \label{formula:LIMA}
\end{eqnarray}
where $N_{\mathrm{on}}$ and $N_{\mathrm{off}}$ represent the counts in the source and background regions, respectively. 
The source regions are defined as circles with a radius of $5^{\prime\prime}$ centered on the source positions, 
while the background regions are concentric annuli with an inner radius of $8^{\prime\prime}$ and an outer radius of $27^{\prime\prime}$. 
The parameter $\alpha$ is the ratio of the source region area to the background region area, which equals $0.038$ for our choice of region geometry. 
By applying a significance threshold of $S > 3$, we obtained a final sample of 79, 22, and 94 hotspots in the NW, SW, and SE regions, respectively. 
Representative images of the detected X-ray hotspots in each region are presented in Supplementary Fig.~\ref{sfig:closeup}. 
We note that \texttt{wavdetect} candidates that do not satisfy $S > 3$ are excluded from the final sample and thus are not marked in Fig.~\ref{fig:NW,SW,SE image}.

\subsection*{Spectral Analysis}\label{section:Spectra}
We performed spectral analysis of the hotspots detected in RX~J1713.7$-$3946 to characterize their emission mechanisms and physical properties. 
In their spectral analysis of the hotspots in the NW region, Higurashi et al.\cite{Higurashi2020} extracted the source spectra from circular regions with  a radius of $5^{\prime\prime}$ and 
subtracted the background accumulated from concentric annular regions with an inner radius of $8^{\prime\prime}$ and an outer radius of $27^{\prime\prime}$. 
However, simply subtracting backgrounds from nearby regions would lead to large statistical uncertainties in determination of spectral parameters, especially for faint hotspots. 
We thus employ a simultaneous fitting method, where the source and background models are determined by jointly fitting the spectra from both the source and background regions. 

In our spectral analysis, the emission within a given source region is assumed to be composed of the following three components: (i) the hotspot emission, (ii) the X-ray background, which includes the diffuse SNR emission, the Galactic ridge X-ray emission, and the cosmic X-ray background, and (iii) the non-X-ray background (NXB). 
Correspondingly, spectra extracted from a background region are modeled to include only of the components (ii) and (iii). 
Given the proximity of the source and background regions, we assume that these two background components share identical spectral shapes and intensities (normalized by the region area)
between the two regions. 

We performed simultaneous fits to the source and background spectra to self-consistently determine the spectral parameters for all components. 
The spectral fitting was carried out using XSPEC\cite{XSPEC1996} version 12.13.1e. We employed the Cash statistic\cite{Cash1979}, implemented as \texttt{cstat} in XSPEC, as the fit statistic for all spectral analyses.
For components (i) and (ii), we employed an absorbed power-law model, $\texttt{TBabs} \times \texttt{pegpwrlw}$, where \texttt{TBabs} represents the Tuebingen-Boulder ISM absorption model\cite{Wilms2000} and \texttt{pegpwrlw} denotes a power-law model.   
We linked the equivalent hydrogen column density ($N_\mathrm{H}$) between components (i) and (ii). 
The NXB component was modeled using the \texttt{mkacispback} tool\cite{Suzuki2021}.

\subsection*{Temporal Variability Searches}\label{section:Variability}
We systematically searched for temporal variability of the X-ray hotspots by using light curves with fluxes derived from the spectral analysis for each epoch. 
For the NW region, the datasets from the observations in 2005 (ObsID 5560) and 2006 (ObsID 6370) were combined due to the relatively short interval between them. 
Similarly, datasets from the three observations in 2009 (ObsIDs 10090, 10091, and 10092) were grouped into a single epoch. 
Spectral fittings for each hotspot at each epoch were performed with the parameters of the X-ray components fixed to the best-fit values from the time-averaged spectra, except for the normalizations of the hotspot and X-ray background components. 
The NXB component was modeled independently for each epoch. 
The normalization of the X-ray background component was allowed to vary, as the diffuse SNR emission itself can be variable, as reported previously\cite{Uchiyama2007}. 
Finally, the light curves for each hotspot were fitted with a constant function, and the significance of the temporal variability was evaluated based on the null hypothesis probabilities.

Supplementary Fig.~\ref{sfig:variable} shows the locations of the hotspots that exhibited time variability with significances of $> 3\sigma$ and $> 2\sigma$. 
We detected significant time variability for 18 hotspots, with only one of them located in the ``Outer'' sub-regions. 
The physical implication of this result follows from the short synchrotron cooling time of the emitting electrons. 
For the clump field $B_\mathrm{clump} \sim 1~\mathrm{mG}$, electrons radiating in the X-ray band cool on a timescale of $\sim 1~\mathrm{yr}$, consistent with the rapid variability of the downstream filaments\cite{Uchiyama2007}. 
The hotspot brightness therefore tracks the instantaneous rate at which secondary electrons are supplied. 
If the proton flux diffusing into the upstream were intermittent, the upstream hotspots would flicker on a timescale of a few years. 
The observed constancy of the upstream hotspots over the decade-long baseline---with only one time-variable hotspot in an ``Outer'' sub-region---thus provides direct evidence that the supply of upstream protons, and hence the diffusion of accelerated protons into the ambient medium, proceeds in a steady state.

\subsection*{Upstream Hotspot Flux and Hadronic Scenario}\label{section:Luminosity}
In the secondary-electron scenario, the same $pp$ interactions are responsible for both the synchrotron X-ray emission from the hotspots and $\pi^{0}$-decay gamma rays.  
The flux ratio between the two radiation channels can, therefore, be predicted. 
By summing the upstream hotspot X-ray fluxes in the NW region, the total is estimated to be $\sim 1 \times 10^{-13}~\mathrm{erg}~\mathrm{cm}^{-2}~\mathrm{s}^{-1}$ at 1~keV. 
We also estimated the TeV gamma-ray flux from the ``Outer'' sub-region in the NW. 
We obtained a flux of $\sim 1 \times 10^{-12}~\mathrm{erg}~\mathrm{cm}^{-2}~\mathrm{s}^{-1}$ at 0.1~TeV by scaling the H.E.S.S. spectrum of the whole remnant by the ratio of the sub-region's gamma-ray flux to that of the entire remnant, calculated using the H.E.S.S. image. 
This value should be considered as a rough estimate since the upstream and downstream emissions cannot be well resolved with the point spread function of H.E.S.S. 
These estimates lead to a flux ratio of $F_\mathrm{X}/F_{\gamma} \sim 0.1$, which is broadly consistent with the theoretical model curves calculated across a wide range of parameters\cite{Tsuji2024}.

\subsection*{Estimate of the Upstream Magnetic Field Strength}\label{section:Estimate}
Based on the $l_\mathrm{diff}$ value estimated from the hotspot distribution, we can derive the magnetic field strength in the shock upstream. 
We define the diffusion length $l_\mathrm{diff}$ as the distance ahead of the forward shock at which the smoothed surface density of the upstream hotspots declines to $1/e$ of its value at the shock front (Fig.~\ref{fig:radial_profile}).
The diffusion length can be written as $l_\mathrm{diff} = \kappa/V_\mathrm{sh}$, where $\kappa$ and $V_\mathrm{sh}$ are the diffusion coefficient and the shock velocity, respectively. 
Inoue et al.\cite{inoue2024} provided a practical expression for $l_\mathrm{diff}$ as 
\begin{eqnarray}
  l_{\mathrm{diff}}=0.1\eta \left(\frac{E_{p}}{10~\mathrm{TeV}}\right)\left(\frac{B_{\mathrm{up}}}{10~\mathrm{\mu G}}\right)^{-1}\left(\frac{V_{\mathrm{sh}}}{3000~\mathrm{km~s^{-1}}}\right)^{-1}~\mathrm{pc}\label{formula:l_diff}, 
\end{eqnarray}
where $B_{\mathrm{up}}$, $E_{p}$, and $\eta$ are the upstream magnetic field strength, the proton energy, and the gyrofactor, respectively. 
For the value for $V_{\mathrm{sh}}$, we referred to the proper motion measurements in the literature\cite{TsujiUchiyama2016,Tanaka2020,Acero2017}. 
To derive $B_{\mathrm{up}}$ using Eq.~\ref{formula:l_diff}, we must first determine $E_{p}$ of the parent protons responsible for the X-ray hotspot emission via secondary electron synchrotron radiation. 
The secondary electron spectra peak at 
\begin{eqnarray}
  E_{e} \sim \frac{E_{p}}{20}\label{formula:E_p,E_e}
\end{eqnarray}
in a wide range of $E_{p}$ between 0.1~TeV and 1000~TeV\cite{Kelner2006}. 
The synchrotron photon energy ($\varepsilon_\mathrm{sync}$) and the parent electron energy ($E_{e}$) are related as 
\begin{eqnarray}
  E_{e}&=&7\left(\frac{\varepsilon_{\mathrm{sync}}}{1~\mathrm{keV}}\right)^{1/2}\left(\frac{B_{\mathrm{clump}}}{1~\mathrm{mG}}\right)^{-1/2}~\mathrm{TeV},\label{formula:E_e}
\end{eqnarray}
where $B_{\mathrm{clump}}$ represents the magnetic field strength in the emitting region---in this case, the dense molecular clumps. 
We assume $B_{\mathrm{clump}} = 1~\mathrm{mG}$ based on the empirical relationship between magnetic field strength and cloud density\cite{Crutcher1999}: 
\begin{eqnarray}
  B_{\mathrm{clump}} \sim \left(\frac{n_\mathrm{clump}}{10^6~\mathrm{cm}^{-3}} \right)^{1/2}~\mathrm{mG},
\end{eqnarray}
where we assume a typical clump number density of $n_\mathrm{clump} = 10^6~\mathrm{cm}^{-3}$. 
Substituting Eqs.~\ref{formula:E_e} and \ref{formula:E_p,E_e} into Eq. \ref{formula:l_diff}, we finally obtain 
a direct constraint on the upstream magnetic field strength as follows: 
\begin{eqnarray}
  B_{\mathrm{up}} = 14\,\eta \left(\frac{\varepsilon_{\mathrm{sync}}}{1~\mathrm{keV}}\right)^{1/2}\left(\frac{B_{\mathrm{clump}}}{1~\mathrm{mG}}\right)^{-1/2}
   \left(\frac{V_{\mathrm{sh}}}{3000~\mathrm{km~s^{-1}}}\right)^{-1}\left(\frac{l_{\mathrm{diff}}}{1~\mathrm{pc}}\right)^{-1}~\mathrm{\mu G}.\label{formula:Bdiff}
\end{eqnarray}
This estimate represents a spatial average of the upstream magnetic field since a uniform field structure is assumed in the calculation.

In Eq.~\ref{formula:l_diff}, the gyrofactor is defined as $\eta \equiv (B_\mathrm{up}/\delta B)^{2} \geq 1$, where $\eta = 1$ corresponds to the Bohm limit\cite{inoue2024}. 
The value $B_\mathrm{up} \sim 10~\mathrm{\mu G}$ quoted in the main text corresponds to the Bohm limit, $\eta = 1$, throughout the upstream region. 
In reality, the gyrofactor can vary with distance from the shock; it likely increases as a function of distance. 
Since the observationally determined $l_\mathrm{diff}$ constrains the product $\eta^{-1} B_\mathrm{up}$, 
our estimate of $B_\mathrm{up}$ directly depends on the assumption regarding $\eta$. 
This, however, does not affect the estimate of the proton maximum energy ($E_\mathrm{max}$) since it is also proportional to the same product as 
\begin{eqnarray}
  E_\mathrm{max} =  96 \, \eta^{-1} \left(\frac{B_\mathrm{up}}{14~\mu\mathrm{G}}\right) \left(\frac{V_{\mathrm{sh}}}{3000~\mathrm{km~s^{-1}}}\right)^{2} \left(\frac{t_\mathrm{age}}{1600~\mathrm{yr}}\right)~\mathrm{TeV},\label{formula:Emax}
\end{eqnarray}
where $t_\mathrm{age}$ is the age of the SNR. 
Therefore, our conclusion that RX~J1713.7$-$3946 is not a PeVatron is independent of the assumed value of $\eta$.

\subsection*{Leptonic Scenario}\label{section:leptonic}
We cannot totally rule out the possibility that the emitting particles in the upstream hotspots are primary electrons accelerated at the forward shock. 
Even in this case, we can constrain $B_\mathrm{up}$ in the same manner as in the secondary electron scenario described above. 
From Eq.~\ref{formula:E_e}, the energy of the radiating electrons is calculated as $E_e = 7~\mathrm{TeV}$.  
Since Eq.~\ref{formula:l_diff} can also be applied to electrons, we can solve the equation for $B_\mathrm{up}$ after substituting  $E_e = 7~\mathrm{TeV}$ into it. 
This yields an upstream magnetic field strength of $B_{\mathrm{up}} \sim 0.7~\mathrm{\mu G}$. 
Relativistic electrons efficiently lose energy through synchrotron radiation; therefore, electron diffusion can, in principle, also be limited by this cooling process. 
Under such a weak magnetic field, the synchrotron cooling timescale is calculated to be $t_\mathrm{sync} \sim 10^{5}~\mathrm{yr}$, making the cooling effect negligible compared to the upstream residence time of the electrons, $t_\mathrm{res} = l_\mathrm{diff}/V_\mathrm{sh} \sim 300~\mathrm{yr}$. 
With this low value of $B_\mathrm{up}$, we can conclude that protons cannot be accelerated up to PeV energies in the leptonic scenario as well.

\clearpage

\clearpage
\section*{Data Availability}\label{sec:data_availability}
The data analysed in this study are archival Chandra ACIS event files. They are publicly available, without restriction, from the Chandra Data Archive (CDA) provided by the Chandra X-ray Center (CXC) at \url{https://cxc.harvard.edu/cda/}, where they are permanently stored. The specific Observation IDs (ObsIDs) used for our analysis are summarized in Supplementary Table~\ref{sTable:Chandra-Obs}. The TeV gamma-ray profiles plotted in Fig.~\ref{fig:radial_profile} were taken from ref.\cite{HESSCollaboration2018}. The data products generated in this study (the hotspot catalogue, the extracted spectra and the spectral-fit results) are stored on institutional servers at Konan University and are available from the corresponding author (Y.K.). No restrictions apply to their access or use.

\section*{Code Availability}\label{sec:code_availability}
For our data reduction and analysis, we used the public software Chandra Interactive Analysis of Observations (CIAO)\cite{CIAO2006}, developed and maintained by the CXC and available at \url{https://cxc.harvard.edu/ciao/}. 
We performed the X-ray spectral fitting with XSPEC\cite{XSPEC1996}, which is publicly available at \url{https://heasarc.gsfc.nasa.gov/docs/software/xspec/} via the High Energy Astrophysics Science Archive Research Center (HEASARC) at NASA's Goddard Space Flight Center. 

\section*{Competing interests}\label{sec:competing}
The authors declare no competing interests.

\section*{Acknowledgements}\label{sec:acknowledge}
This research has made use of data obtained from the Chandra Data Archive provided by the Chandra X-ray Center (CXC). 
We thank Hiroyuki Uchida and Shiu-Hang Lee for fruitful discussions.

\section*{Funding}
This work was supported by JSPS KAKENHI Grant Numbers JP19H01936, JP21H04493, JP23H00129, JP23H01211, JP24K17093, and JP26K00714. 

\section*{Author contributions}\label{sec:author_cont}
Y.K. conducted the data analysis under the supervision of T.T. 
H.S. contributed to the background modeling in the spectral analysis. 
T.I. led the theoretical interpretation of the observational results. 
The manuscript was written mainly by Y.K. and T.T., and was reviewed and commented on by all authors.

\setcounter{figure}{0}
\renewcommand{\tablename}{Supplementary Table}
\renewcommand{\figurename}{Supplementary Figure}

\begin{table}
  \caption{Chandra Observation Log.} \label{table:Chandra Observations of RX J1713.7--3946 NWSWSE}
\begin{tabular}{ccccc}
\hline
ObsID & Target Name & Start Date & (RA, Dec)$_{\mathrm{J2000.0}}$ & Exposure (ks) \\
\hline
736   & RX~J1713.7$-$3946 NW & 25 July 2000 & (17:11:48.00, $-$39:34:00.00) & 29.6 \\
5560  & RX~J1713.7$-$3946 NW & 09 July 2005 & (17:11:45.50, $-$39:33:23.20) & 29.0 \\
6370  & RX~J1713.7$-$3946 NW & 03 May 2006  & (17:11:45.50, $-$39:33:23.20) & 29.8 \\
10090 & RX~J1713.7$-$3946 NW & 30 Jan 2009  & (17:11:45.50, $-$39:33:23.20) & 28.4 \\
10091 & RX~J1713.7$-$3946 NW & 16 May 2009  & (17:11:45.50, $-$39:33:23.20) & 29.6 \\
10092 & RX~J1713.7$-$3946 NW & 10 Sept 2009 & (17:16:03.00, $-$39:57:17.00) & 29.2 \\
12671 & RX~J1713.7$-$3946 NW & 01 July 2011 & (17:11:45.50, $-$39:33:23.20) & 89.9 \\
\hline
5561  & RX~J1713.7$-$3946 SW & 09 July 2005 & (17:11:49.99, $-$39:54:29.88) & 29.0 \\
21339 & RX~J1713.7$-$3946 SW & 17 May 2020  & (17:11:50.00, $-$39:54:29.90) & 29.7 \\
\hline
737   & RX~J1713.7$-$3946 SE & 25 July 2000 & (17:15:16.80, $-$39:57:36.00) & 38.5 \\
10697 & RX~J1713.7$-$3946 SE & 15 May 2009  & (17:11:45.50, $-$39:33:23.20) & 57.4 \\
\hline
6721  & CTB~37A              & 07 Oct 2006  & (17:14:35.08, $-$38:31:24.06) & 19.9 \\
14586 & IC~4637              & 07 Mar 2014  & (17:05:10.05, $-$40:53:08.04) & 29.6 \\
\hline
\end{tabular}
\label{sTable:Chandra-Obs}
 \end{table}

\begin{figure}[p]
\centering
\includegraphics[width=\linewidth]{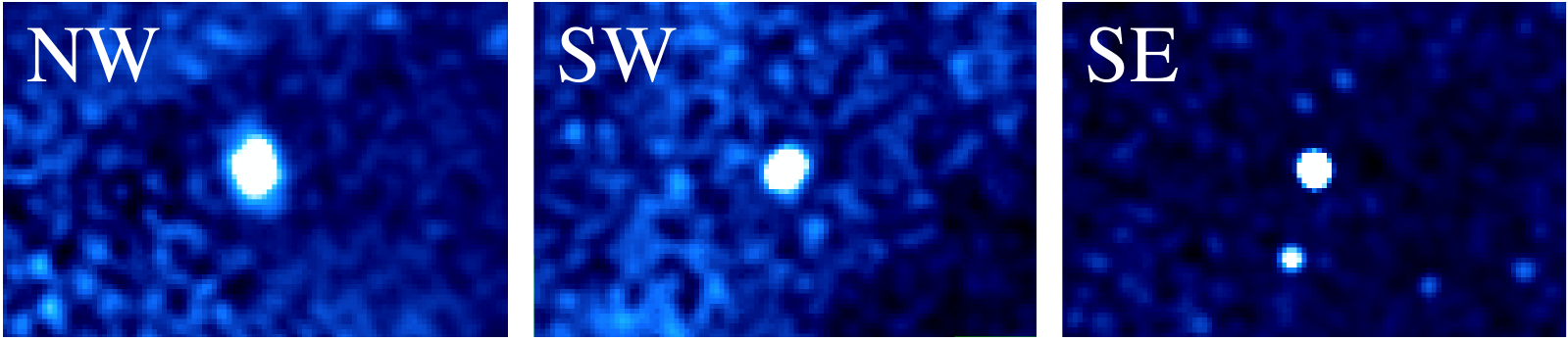}
\caption{Close-up views of representative X-ray hotspots. Each panel displays
a 0.5--7.0\,keV image of a $150'' \times 100''$ area in the NW, SW, and SE
regions, centered on a detected hotspot.}
\label{sfig:closeup}
\end{figure}

\begin{figure}[p]
\centering
\includegraphics[width=\linewidth]{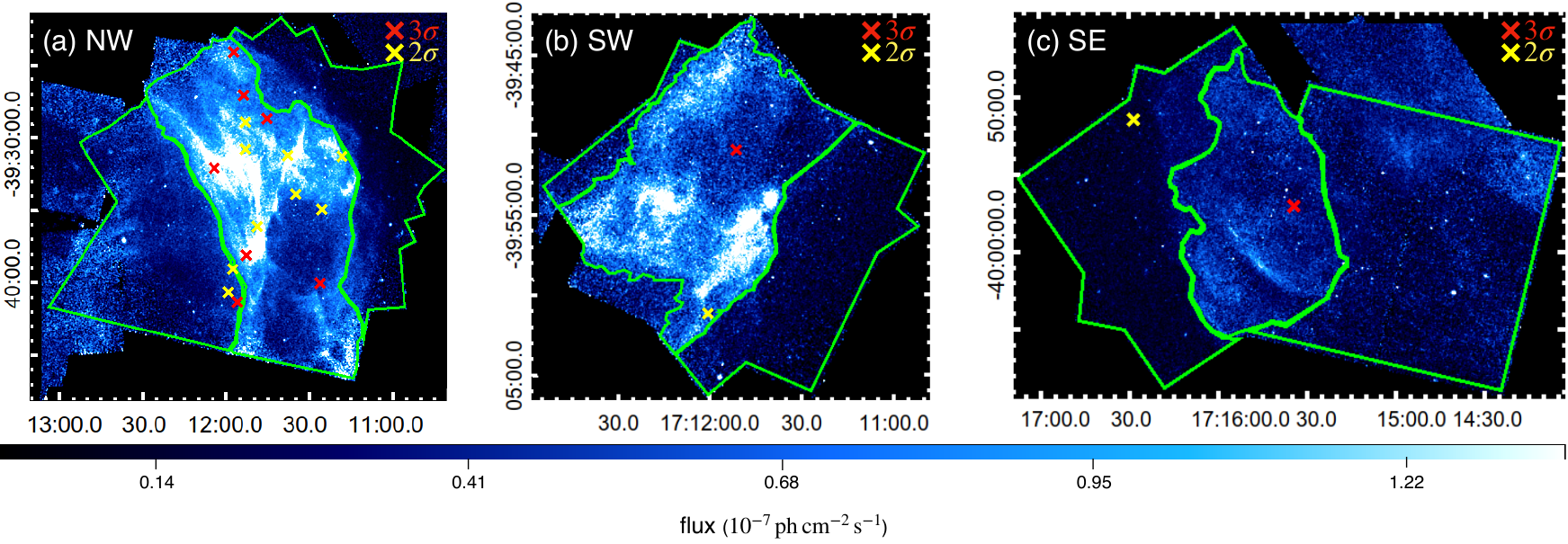}
\caption{Locations of X-ray hotspots with significant time variability.
Red and yellow crosses, overlaid on the Chandra X-ray images of the (a) NW,
(b) SW, and (c) SE regions, denote the locations of the hotspots that
exhibited time variability with significances of $>3\sigma$ and $>2\sigma$,
respectively. The spatial distribution shows that most of the time-variable
hotspots are located within the SNR shell, with only one hotspot in the SE
``Outer'' sub-region.}
\label{sfig:variable}
\end{figure}

\end{document}